\documentclass[sigconf]{acmart}

\usepackage{booktabs} 

\usepackage[utf8]{inputenc}
\usepackage[english]{babel}
\usepackage{amsmath}
\usepackage{amsfonts}
\usepackage{amssymb}
\usepackage{graphicx}
\usepackage{bm} 
\usepackage{enumitem} 
\usepackage{caption}
\usepackage{subfig}

\usepackage{listings} 
\usepackage{color}    
\definecolor{mygreen}{rgb}{0,0.6,0}
\definecolor{mygray}{rgb}{0.5,0.5,0.5}
\definecolor{mymauve}{rgb}{0.58,0,0.82}

\usepackage{wrapfig} 

\makeatletter

\hypersetup{
pdfauthor={\@author},
pdftitle={\@title}
}

\begin{document}

\title{x3ogre: connecting X3D to a state of the art rendering engine}

\author{Pavel Rojtberg}
\email{pavel.rojtberg@igd.fraunhofer.de}
\affiliation{%
  \institution{Fraunhofer IGD}}
\author{Benjamin Audenrith}
\email{benjamin.audenrith@igd.fraunhofer.de}
\affiliation{%
  \institution{Fraunhofer IGD}}

\keywords{X3D, webgl, materials, geometry}

\lstset{
commentstyle=\color{mygreen},
stringstyle=\color{mymauve},
showspaces=false,
captionpos=b,
showstringspaces=false,
tabsize=2,
frame=none,
basicstyle=\footnotesize,
caption=\relax
}

\acmYear{2017}
\setcopyright{acmlicensed}
\acmConference{Web3D '17}{June 05-07, 2017}{Brisbane, QLD,
Australia}\acmPrice{15.00}\acmDOI{http://dx.doi.org/10.1145/3055624.3075949}
\acmISBN{978-1-4503-4955-0/17/06}

\begin{abstract}
We connect X3D to the state of the art OGRE renderer using our prototypical x3ogre implementation. At this we perform a comparison of both on a conceptual level, highlighting similarities and differences.
Our implementation allows swapping X3D concepts for OGRE concepts and vice versa. We take advantage of this to analyse current shortcomings in X3D and propose X3D extensions to overcome those.
\end{abstract}

%
%

\ccsdesc[500]{Computing methodologies~Graphics file formats}

\maketitle


\section{Introduction}
X3D \cite{daly2007x3d} is an open standard for 3D graphics with precisely defined semantics. Scenes stored in the X3D format can be parsed using standard XML parsers and the files are usually self-contained which makes X3D a good choice for interchange.
However the standardization process causes that new rendering techniques and concepts appear in X3D with a considerable delay. The available X3D based rendering engines like X3DOM \cite{Behr:2009:XDH:1559764.1559784} or InstantReality \cite{avalon} therefore offer custom extensions to overcome this. Yet those are only scarcely used as they impede interchange.

This paper therefore takes a different approach and instead connects X3D to an existing state of the art rendering engine. This allows using the file formats of the underlying renderer where X3D falls short. As these formats neither are standardized nor have to provide legacy compatibility, they can evolve faster and at the same time are better optimized for rendering.
In comparison with creating an X3D extension this approach is more flexible as it allows replacing even core X3D concepts.
This flexibility in turn gives better insights on how to improve X3D itself.

Here, we focus on the presentation aspect of X3D as it is arguably the most important part of a 3D file format. When using the X3D DOM profile \cite{Behr:2009:XDH:1559764.1559784} most application logic is outside of the X3D format and recent extensions (e.g.\ \cite{schwenk2012commonsurfaceshader} and \cite{sturm2016unified}) specifically target material representation.
Therefore, we neglect the Scripting and Sensor parts of X3D and concentrate the Rendering and Geometry components.

For the underlying renderer we chose the Object-Oriented Graphics Rendering Engine (OGRE) \cite{ogre3d} . While other rendering engines like Unreal 4 and Unity recently gained more popularity, OGRE is available royalty-free under an permissive open-source license making it a better fit for research as well as allowing deeper inspection.

\begin{figure}
\subfloat[Per-pixel lighting] {
\includegraphics[width=0.23\textwidth]{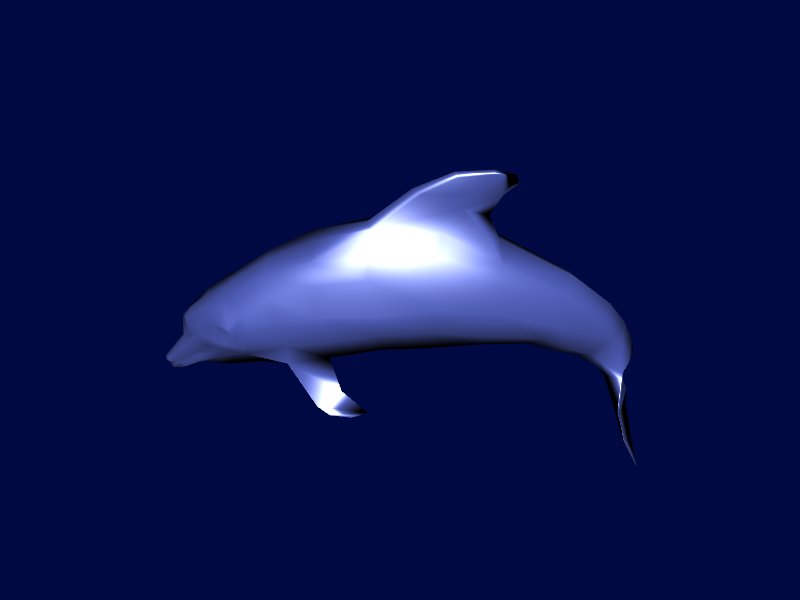}
}
\subfloat[Night vision compositor effect] {
\label{fig:nightvis}
\includegraphics[width=0.23\textwidth]{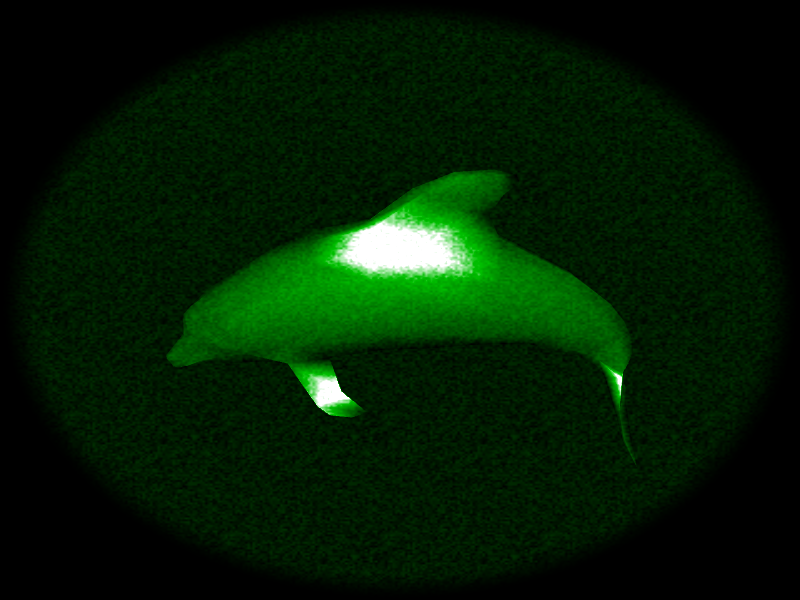}
}
\caption{The flipper.x3d example showcasing some of the OGRE features}
\label{fig:dist_effect}
\end{figure}

OGRE is not bound to a specific rendering API like OpenGL or DirectX, but rather provides high level concepts that map to any of those. It is being developed since 1999 and was used in AAA games like Torchlight 1/2 as well as in industrial training applications. 
Therefore the rendering concepts are mature and proven --- while having been developed independently to X3D. This makes a comparison especially interesting.

The comparison is performed using our prototypical implementation called "x3ogre", which allows loading X3D scenes in OGRE as well as using OGRE concepts inside X3D. 
Besides the comparison, one use-case of the prototype is to visually enhance existing X3D scenes without requiring invasive changes. 


By utilizing Emscripten \cite{zakai2011emscripten} OGRE based applications can be deployed on the Web\footnote{\url{https://ogrecave.github.io/ogre/emscripten/}}. Internally a GLES3 \cite{opengles} based renderer is used, which supports WebGL \cite{webgl} while also providing forward compatibility with WebGL2.
Therefore the results are not limited to the C/ C++ ecosystem, but can also be transferred to the web context.

This paper is structured as follows: in section \ref{sec:x3d2ogre} the mapping of the X3D concepts to the corresponding OGRE concepts is discussed, while section \ref{sec:ogre2x3d} takes the reverse way, describing the integration of OGRE concepts in X3D.
Based on the preceding discussions we then propose several X3D extensions in section \ref{sec:x3dext}.
Finally we conclude with section \ref{sec:conclusion} giving a summary of our results and discussing the limitations and future directions.

\section{Transferring X3D concepts to OGRE}
\label{sec:x3d2ogre}
As the first step we identify the concepts corresponding to X3D in OGRE. Here we focus on a subset of the X3D interchange profile \cite{x3dpart1} which roughly corresponds the X3D DOM profile introduced by \cite{Behr:2009:XDH:1559764.1559784}.

We will briefly discuss the high level objects and then focus on the compound geometry and material objects in more detail.

The correspondences for the core X3D objects are:
\begin{itemize}
\item the X3D Scene node corresponds to the \textit{SceneManager}. In X3D only one Scene node per file is allowed, while OGRE supports several active SceneManager instances. With only one of them being rendered by a specific Camera.
\item the X3D Transform node corresponds to two nested \textit{SceneNode}s. SceneNodes only store a translation vector, an orientation quaternion and a scaling vector. Therefore two SceneNodes are needed to support the \textit{center} property of a Transform. The X3D \textit{scaleOrientation} property can only be implemented for multiples of $90^\circ$ as there is no shearing component in OGRE.

Note that OGRE does not have a native serialization format for SceneNodes and thus benefits by using X3D here.
\item the X3D Geometry node corresponds to the \textit{Mesh} object. As in X3D this is a compound object. It can be serialized either as XML or in a binary file format.
\item the X3D Appearance node corresponds to the \textit{Material} object. Again this is a compound object. It is serialized in a custom file format resembling the VRML97 encoding.
\end{itemize}

Additionally we identified the following correspondences needed to support animations. These are given for completeness as the concepts mostly map one-to-one:
\begin{itemize}[noitemsep]
\item the X3D TimeSensor maps to \textit{AccumulateControllerFunction}.
\item the X3D ScalarInterpolator maps to \textit{LinearControllerFunction}.
\item the X3D CoordinateInterpolator maps to \textit{VertexAnimationTrack}.
\item the X3D PositionInterpolator and OrientationInterpolators correspond to \textit{NodeAnimationTrack}.
\end{itemize}

Note that for supporting animations in X3D one does not need to implement the full X3D event model. Instead it is sufficient to rewrite specific ROUTE statements in a compositional way as shown in listing \ref{route}.
\begin{lstlisting}[label=route,language=XML, caption=Compositional ROUTE implementation]
<TimeSensor DEF="time" />
<PositionInterpolator DEF="move" />
<ROUTE fromNode="time" fromField="fraction_changed" 
       toNode="move" toField="set_fraction" />
<!-- is transformed into -->
<PositionInterpolator DEF="move">
  <TimeSensor USE="time" />
</PositionInterpolator>
\end{lstlisting}

This approach requires allowing specific source nodes as children of the target nodes and is therefore not compatible with general ROUTE statements. However it enabled the correct loading of the existing X3D scenes we tried.

\subsection{Geometry}
OGRE uses the binary .mesh file format which can be transparently converted from and to XML for inspection and debugging.

Listing \ref{x3dgeom} shows a typical X3D geometry definition which translates to the OGRE XML format in listing \ref{ogregeom}.
\begin{lstlisting}[label=x3dgeom,language=XML, caption=Sample X3D Shape definition]
<Shape>
  <Appearance USE='Example' />
  <IndexedTriangleSet coordIndex="0 1 2 [...]">
    <Coordinate point="0 0 0 [...]" />
    <Normal vector="0 0 0 [...]" />
  </IndexedTriangleSet>
</Shape>
\end{lstlisting}
The Shape node corresponds to a submesh node and the IndexedTriangleSet node corresponds to geometry node. Note that OGRE allows interleaved storage of the position and normal vertex attributes while X3D does not. This is useful, as OGRE also stores additional vertex attributes like bone assignments for skeletal animation (HAnimSegment node in X3D) in the .mesh file.
\begin{lstlisting}[label=ogregeom,language=XML, caption=Sample OGRE XML Mesh definition]
<submesh material="Example"
         usesharedvertices="false"
         operationtype="triangle_list">
  <faces count="815">
    <face v1="1" v2="2" v3="3" />
    [...]
  </faces>
  <geometry vertexcount="531">
    <vertexbuffer positions="true" normals="true">
      <vertex>
        <position x="0" y="0" z="0"/>
        <normal x="0" y="0" z="0"/>
      </vertex>
      [...]
    </vertexbuffer>
  </geometry>
</submesh>
\end{lstlisting}
Listing \ref{ogregeom} shows a single submesh definition, but generally a .mesh file stores multiple submeshes. This allows the definition multi-material meshes, where each submesh is rendered with one material.
In X3D one has to use multiple Shapes and then use a Group node for linking them.

Furthermore the submeshes can reference the same vertices to avoid data duplication. This representation directly maps to the low-level graphics APIs. (In OpenGL: \textit{glDrawRangeElements} and \textit{glBufferData}).

For X3D this concept can be reproduced by DEF/ USE of the Coordinate node --- yet due to the complexity of the system it is not obvious whether X3D viewers will share the memory between the corresponding Shapes.

In OGRE, used material can only be referenced in the .mesh file while in X3D it is typically defined inline with the geometry data. This separation is similar to the concept in glTF \cite{robinet2013gltf} or the X3D BinaryGeometry node \cite{behr2012using}.
Mesh files do not specify any compression. However OGRE assets are usually distributed in zipped packs containing geometry, materials and textures which offer compression on a higher level.


\subsection{Materials}

OGRE uses the custom .material script format for material definition which resembles the classic VRML encoding. For comparability we will use the VRML encoding for the following X3D material examples.
\begin{lstlisting}[label=x3dmat,language=C++, caption=Example X3D Appearance definition]
DEF Example Appearance {
  material Material {
    ambientIntensity 0.508497
    diffuseColor 0.337255 0.4 0.788235
    specularColor 1 1 1
} }
\end{lstlisting}
The X3D material in listing \ref{x3dmat} translates to the OGRE material given in listing \ref{ogremat}.
\begin{lstlisting}[label=ogremat,language=C++, caption=According OGRE Material definition]
material Example {
  technique {
    pass {
      ambient 0.508497 0.508497 0.508497
      diffuse 0.337255 0.4 0.788235
      specular 1.0 1.0 1.0 25
} } }
\end{lstlisting}
OGRE materials support multiple techniques which are again composed of several rendering passes. The technique range allows picking the appropriate one at runtime based on hardware support, LOD level etc., while defining multiple passes can be useful for advanced rendering techniques like rendering hair.


Both  OGRE and X3D material definitions reflect the simple Blinn-Phong shading model \cite{Blinn1977BlinnPhongShading}. However state of the art rendering usually involves more sophisticated lighting models like the Cook-Torrance microfacet reflection model \cite{Cook1981CookTorranceModel} --- optionally combined with normal mapping and deferred shading.

This requires a more flexible material definition. \cite{schwenk2010modern} therefore introduced the X3D CommonSurfaceShader node that used the uber-shader \cite{hargreaves2005generating} approach. While offering more flexibility then the traditional materials, the monolithic nature requires the change of existing materials whenever a new rendering technique must be incorporated; the CommonSurfaceShader had to be updated to incorporate Physically Based Shading (PBS) \cite{schwenk2012commonsurfaceshader}.

In contrast OGRE provides the high level material system (HLMS) for defining custom materials. This system builds around the idea of passing opaque properties to a named template shader.
\begin{lstlisting}[label=x3dpbs,language=XML, caption=Physically based material in X3D]
<Appearance DEF="Example">
  <PhysicalMaterial albedoFactor="0.22 0.3 0.5"
    roughnessFactor="0.4",
    metallicFactor="0.76" />
</Appearance>
\end{lstlisting}
For instance the PhysicalMaterial node of \cite{sturm2016unified} (listing \ref{x3dpbs}) translates to the following HLMS material in OGRE.
\begin{lstlisting}[label=ogrepbs,language=C++, caption=Physically based material in OGRE]
hlms Example PBS {
  diffuse   1 1 1
  specular  1 1 1
  roughness 0.4
  fresnel   1.3
}
\end{lstlisting}
However while \cite{sturm2016unified} rely on a predefined Shader, OGRE just forwards the given parameters to a shader named "PBS" (HLSL on DirectX , GLSL on OpenGL). This allows users to define custom materials with arbitrary parameters. Figure \ref{fig:pbs} shows a grid of spheres rendered with custom PBS shading while the ground plane is being rendered using the classical Blinn-Phong shading.

\section{Connecting OGRE to X3D}
\label{sec:ogre2x3d}
In Section \ref{sec:x3d2ogre} we identified the X3D concepts inside OGRE which allowed loading X3D files. This section on the other hand will focus on bringing the OGRE concepts to X3D.
First we describe the connection for concepts which exist in both X3D and in OGRE, where the OGRE counterparts usually offer more flexibility. Then we will describe the mapping of the OGRE compositor system for which X3D has no counterpart.

\subsection{Connection of concepts}

The general approach taken here is to redirect existing X3D concepts to their OGRE counterparts as identified in section \ref{sec:x3d2ogre}. In contrast to creating new X3D nodes, this simplifies the implementation and allows to first evaluate the benefits before introducing new concepts in X3D.
For instance to use the interleaved vertex attribute storage in X3D we just redirect the Geometry node to an OGRE .mesh file and override the material by the X3D appearance. Listing \ref{refogre} shows the different redirection options we support in x3ogre.
\begin{lstlisting}[label=refogre,language=XML, caption=Using OGRE formats inside X3D]
<!-- OGRE defined mesh, may reference X3D materials -->
<Shape USE="Sinbad.mesh" />
<!-- OGRE defined material in a X3D Shape -->
<Shape>
  <Appearance USE='Ogre/ExampleMaterial' />
  <IndexedTriangleSet>[...]</IndexedTriangleSet>
</Shape>
<!-- OGRE defined geometry in a X3D Shape -->
<Shape>
  <Appearance>[...]</Appearance>
  <Geometry USE="Sindbad.mesh" />
</Shape>
\end{lstlisting}
Both, the X3D DEF/ USE system and the OGRE resource system rely on strings for identifying resources. The implementation therefore is straightforward and we can maintain compatibility with existing X3D files by only using OGRE resources if the USE lookup inside of the current X3D file fails.

Compared to the ExternalGeometry node proposed by \cite{limper2014src} this approach is easier to implement and does not requires invasive changes to existing X3D files. However it implicitly relies on an externally defined resource pool and  overriding individual mesh properties (e.g.\ color) is not possible. 

\subsection{Explicit compositing}
A compositing system allows specifying full screen effects to be applied to the image after a scene has been rendered --- similar to layer effects in an image editing software. A simple example would be desaturation to create a black \& white image without having to change the scene materials.
Figure \ref{fig:nightvis} shows a more complex effect that involves adding noise and vignetting at the image borders.

X3D does not have an explicit compositing concept. Post processing is usually implemented by redirecting the rendering to a RenderedTexture which is then rendered using a custom Appearance on a full-screen quad.
Combining multiple layers is only possible using the MultiTexture node. However it only offers a fixed set of blending modes that correspond to an OpenGL 1.2 extension \cite{texenvcomb}.
Chaining post processing effects is not feasible as there is no mechanism in X3D to specify in which order RenderedTextures should be processed.

\begin{figure}
\subfloat[PBS material system] {
\includegraphics[width=0.23\textwidth]{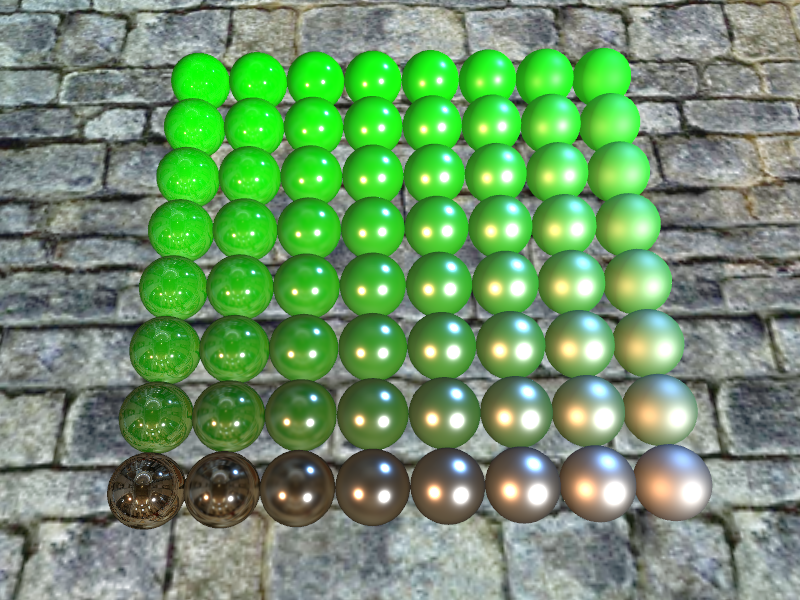}
\label{fig:pbs}
}
\subfloat[Compositor chaining] {
\includegraphics[width=0.23\textwidth]{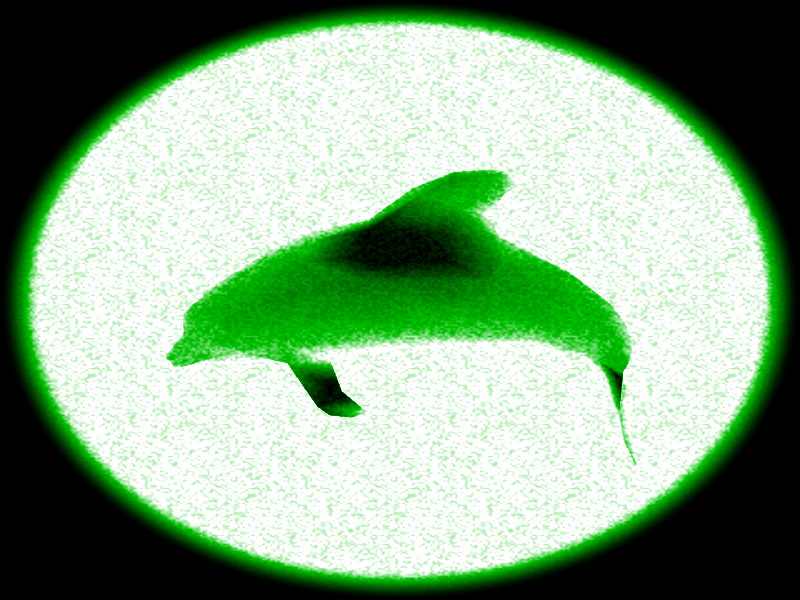}
\label{fig:compchain}
}
\caption{Advanced rendering concepts}
\end{figure}

OGRE on the other hand uses the explicit .compositor file format (see listing \ref{ogrecomp}) that resemble the .material format but instead specifies the rendering and routing of full-screen render targets.
\begin{lstlisting}[label=ogrecomp,language=C++, caption=A simple OGRE compositor effect]
compositor "Night Vision" {
  technique {
    // Temporary texture(s)
    texture rt0 target_width target_height PF_A8R8G8B8
    target rt0 {
      // Render output from previous compositor
      // or original scene
      input previous
    }
    target_output {
      input none
      // Draw a fullscreen quad..
      pass render_quad {
        // ..using the night vision shader
        material Ogre/Compositor/NightVision
        input 0 rt0
} } } }
\end{lstlisting}
Using this format it is possible to describe simple effects in a more concise way compared to X3D, while also allowing complex effects like pre-pending an invert effect to "Night Vision" (figure \ref{fig:compchain}) or even implementing deferred shading.

To enable compositing effects in X3D we added a new MFString field \textit{compositors} to the Viewpoint node which allows specifying a compositor chain for that specific View. (see Listing \ref{compx3d})

\section{Proposed X3D extensions}
\label{sec:x3dext}

Based on the discussion in the preceding sections we now propose two conceptual extensions to X3D that allow implementing the respective OGRE concepts directly instead of merely referencing them.

The first extension is the explicit notion of compositing by introducing a Compositing System inside X3D. The second extension is the definition of user defined appearances.

\subsection{Compositing System}
Following the OGRE notion of a compositor, we propose adding the following X3D nodes to allow the definition of a compositing effect directly inside a X3D file:

\begin{itemize}
\item \textit{Compositor} for defining a named Compositor and the according scope. For defining the intermediate render layers, we use the RenderTexture extension.
\item \textit{CompositorPass} for explicitly stating the rendering order of RenderTextures. For specifying the shader and referencing input textures, we can use the Appearance node without modifications.
\item \textit{CompositorOutput} for explicitly marking the sink of a compositor graph. While one could use a special target on a CompositorPass, this makes automated error checking easier.
\end{itemize}

\begin{lstlisting}[label=compx3d,language=XML, caption=Sample usage of proposed Compositor node]
<Compositor DEF="GaussBlur">
	<RenderedTexture DEF="rt0" />
	<RenderedTexture DEF="rt1" />
	<CompositorPass target="rt0" input="none" render="SCENE"/>
	<CompositorPass target="rt1" input="none" render="QUAD">
	  <Appearance>
        <ComposedShader USE="BlurVertical" />
        <RenderedTexture USE="rt0" />
	  </Appearance>
	</CompositorPass>
	<CompositorOutput input="none" render="QUAD">
	  <Appearance>
        <ComposedShader USE="BlurHorizontal" />
        <RenderedTexture USE="rt1" />
	  </Appearance>
	</CompositorOutput>
</Compositor>
<Viewpoint compositors="GaussBlur" />
\end{lstlisting}

Listing \ref{compx3d} shows a sample usage of the above nodes. The implemented effect is a separated Gaussian Blur filter which requires two render passes to be executed in the correct order.

\subsection{User defined Appearance}

Bringing user defined Appearances to X3D eases using specialized rendering techniques and allows bringing together the proposed Material extensions \cite{schwenk2012commonsurfaceshader} \cite{sturm2016unified} using an unified concept.
\begin{lstlisting}[label=customx3d,language=XML, caption=User defined Appearance in X3D]
<ComposedShader DEF="PBS">
[...]
</ComposedShader>
<CustomAppearance type="PBS">
  <!-- ComposedShader holds the type information -->
  <field name="roughnessFactor" value="0.4" />
  <ImageTexture url="albedo.png" 
                containerField="albedoMap" />
</CustomAppearance>
\end{lstlisting}
To this end we propose the new CustomApperance node that references a named ComposedShader and can be used instead of the classical Appearance node. The CustomApperance node contains any number of Texture and field nodes that are forwarded to the referenced shader.

\section{Conclusion \& Future Work}
\label{sec:conclusion}
We have connected X3D to OGRE in a bidirectional manner allowing X3D scenes to be loaded by OGRE as well as using OGRE resources in X3D scenes. By comparing both on a conceptual level we could identify shortcomings in X3D and propose extensions to overcome those.

However we only extended X3D on a coarse level; one could improve the existing X3D concepts by comparing the implementations in detail. For instance one could improve the Geometry representation in X3D by allowing interleaved storage of vertex attributes and explicit buffer sharing by a notion of submeshes.

Furthermore our implementation, while already runnable on the web, only offers a custom C++ API. To allow using x3ogre as alternative to three.js or X3DOM, a SAI like API must be exported to JavaScript.

Finally it should be evaluated in how far the identified shortcomings also apply to the glTF format.

The implementation presented in this work is available open source at \url{https://github.com/paroj/x3ogre}.

\bibliographystyle{ACM-Reference-Format}
\bibliography{bibliography} 

\end{document}